\documentclass[aip,pop,reprint,numerical,twocolumn,superscriptaddress]{revtex4}
\usepackage{amssymb, amsmath,framed}
\usepackage{color,soul}

\usepackage{bm}
\usepackage[colorlinks=true,linkcolor=blue]{hyperref}
\expandafter\ifx\csname package@font\endcsname\relax\else
 \expandafter\expandafter
 \expandafter\usepackage
 \expandafter\expandafter
 \expandafter{\csname package@font\endcsname}
\fi
\def\bq{\begin{equation}}
\def\eq{\end{equation}}
\def\bqy{\begin{eqnarray}}
\def\eqy{\end{eqnarray}}

\def\de{\delta}

\def\ep{\epsilon}

\def\p{\partial}

\def\rh{\rho}

\def\p{\partial}

\def\calc{\mathcal{C}}

\def\calj{\mathcal{J}}

\def\call{\mathcal{L}}

 \def\q#1#2{q^{\hspace{1pt}#1}_{\,,\hspace{1pt}#2}}
 \def\a#1#2{a^{\hspace{1pt}#1}_{\,,\hspace{1pt}#2}}

\begin{document}

\title{Inertial magnetohydrodynamics}

\author{M.~Lingam}
\email{manasvi@physics.utexas.edu}
\affiliation{Department of Physics and Institute for Fusion Studies, The University of Texas at Austin, Austin, TX 78712, USA}
\author{P.~J.~Morrison}
\email{morrison@physics.utexas.edu}
\affiliation{Department of Physics and Institute for Fusion Studies, The University of Texas at Austin, Austin, TX 78712, USA}

\author{E.~Tassi}
\email{tassi@cpt.univ-mrs.fr}
\affiliation{Aix-Marseille Universit\'e, CNRS \& Centre de Physique Th\'eorique, Campus de Luminy, 13288 Marseille cedex 9, France and Universit\'e de Toulon, CNRS, CPT, UMR 7332, 83957, La Garde, Francé}

\date{\today}

\begin{abstract}
A version of extended magnetohydrodynamics (MHD) that incorporates electron inertia is obtained by constructing an action principle.   Unlike MHD which freezes in magnetic flux, the present theory freezes in an alternative flux related to the electron canonical momentum.   The associated Hamiltonian formulation is derived and reduced models that have previously been used to describe collisionless reconnection are obtained.

\end{abstract}

\maketitle


\section{Introduction} \label{Intro}
Ideal magnetohydrodynamics (MHD), which arose in the early 20th century, ranks amongst the great breakthroughs in plasma physics. The simplicity of MHD has led to its extensive usage in the arenas of fusion, space and astrophysical plasmas; see e.g. \cite{KT73,GP04,K05} and references therein. In addition, MHD is an attractive theory as it is endowed with several geometric properties such as flux freezing and the conservation of magnetic and cross helicities, amongst others. The latter duo, in particular, have attracted much attention owing to their topological properties \cite{M69,BF84} and intimate connections with self-organization and relaxation \cite{W58,T74,T86}. As the advantages of ideal MHD are far too numerous to elaborate, we refer the reader to the aforementioned references. 

However, in the latter half of the 20th century, an increasing awareness arose amongst the plasma community that MHD could serve as an effective theory only in certain regimes. To counteract these limitations, several fluid and kinetic models were developed, of which we shall only focus on the former. Amongst them, the most famous are the two-fluid model \cite{KT73,GP04}, Hall MHD \cite{L60}, electron MHD \cite{RSS91} and extended MHD \cite{S56,L59}. Extended versions of ideal MHD often incorporate additional terms into Ohm's law, but they (mostly) suffer from a common deficiency - a failure to conserve the energy, in the absence of dissipative effects. This fact was first pointed out in \cite{KM14}, who also presented an analysis of the different terms in the extended Ohm's law and their role in energy conservation; also see \cite{pjmTTC14} for a related analysis. Even amongst MHD models, most versions tend to neglect the electron inertia, which can be of considerable interest when their characteristic velocity is much faster than their ionic counterparts. In order to retain electron inertia while simplifying other features, several reduced MHD (RMHD) models have been proposed \cite{OP93,SPK94}. Such models are of considerable interest, as they represent alternative methods of driving reconnection, which in turn is expected to play a crucial role in terrestrial and astrophysical plasmas \cite{ZY09}.

The energy-conserving property of MHD and its invariants are closely linked to its Hamiltonian structure, i.e. it can be shown that such theories can be described via a noncanonical bracket and an appropriate Hamiltonian. The existence of such a Hamiltonian structure for MHD was first recognized in the seminal paper of Morrison \& Greene \cite{MG80}, and was consequently employed in a wide range of contexts in the 1980s \cite{M80,SK82,HK83,MH84}. We refer the reader to \cite{pjm82,S88,M98,pjm05}, and references therein, for comprehensive reviews of the same. The Hamiltonian formalism has advantages that extend far beyond the mere ability to reproduce the equations of motion; it can be used to obtain invariants, equilibria and conduct stability analyses \cite{M98,HMRW85,pjmAP13,pjmTT13}. 

Associated with Hamiltonian structure is the action principle formalism, and this can serve as a starting point for obtaining the noncanonical Poisson bracket.  Such an action principle dates back to Lagrange's pioneering work in the 18th century \cite{L18thcen}, but only relatively recently was such  employed in the context of MHD  \cite{Newcomb62} and,  subsequently,  for incompressible gyroscopic systems \cite{N72,N83}. 

Here, we construct an action by employing  a  method for building actions that was described in  \cite{Mor09}.  An important feature of this method is the {\it Eulerian closure principle} (ECP) that insures  the resulting theory can be written in terms of an Eulerian set of variables.   The   method  has proven to be successful for deriving  compressible  gyroviscous MHD \cite{MLA14}, extended MHD \cite{KLMWW14} and generalized gyroviscous fluids \cite{LM14}.  The method  begins  with an action in terms of a set of  Lagrangian variables, treating the particle trajectory as the sole variable, together with the ECP.   The ECP provides a generalization of,  and indeed a justification of,   constrained variations that were  called  Euler-Poincar\'e variations in \cite{HMR98a,HMR98b,C98}, because of its antecedents in the work of Poincar\'e \cite{Poincare1901}, although the idea was previously elucidated for ideal fluids and MHD plasmas in,  e.g., \cite{FR60,Newcomb62}.   More recently, Euler-Poincar\'e  variations were  employed for gyrokinetic theory in  \cite{brizard00} and gyro fluids in  \cite{brizard10}.

The action formalism also has several other advantages in addition to its ability to reproduce energy conserving models. A second advantage of this approach is that one can obtain reduced or simplified fluid models by performing suitable orderings directly \emph{within} the action. These modifications still preserve energy conservation, which is not guaranteed when an ordering is undertaken at the level of the equations of motion. Thirdly, one can recover the noncanonical bracket and the Hamiltonian described previously; the former via a systematic reduction of the canonical bracket, and the latter via a Legendre transform. As the action principle formulation is intimately linked with the Hamiltonian approach, we refer to them collectively as the Hamiltonian and Action Principle (HAP) formalism, and illustrate its use in the current paper. Lest we lose sight of our goal, we summarize it before proceeding further. In this paper, we shall use the HAP formalism to construct a MHD model with electron inertia. 

The organization of the paper is as follows. In Section \ref{ActPrin}, we introduce the requisite mathematical preliminaries. In Section \ref{IMHDaction}, we introduce a new dynamical variable, construct the action and obtain the equations of motion. In Section \ref{HamForm}, we obtain the noncanonical Hamiltonian formalism via reduction, and demonstrate how it constitutes a generalization of the widely employed Ottaviani-Porcelli model \cite{OP93}. We also briefly comment on possible extensions of this model, and finally conclude in Section \ref{Conc}.

 \section{The action principle and the Lagrangian coordinates}
 \label{ActPrin}
We commence with a very brief synopsis of the Lagrangian approach and the Lagrangian coordinate. For finite dimensional systems, the action is
\bq
S[q] = \int^{t_1}_{t_0} \! dt\,L\left(q,\dot{q},t\right)\,,
\eq
where $L=T-V$ represents the Lagrangian, with the kinetic and potential energies denoted by $T$ and $V$ respectively. Here, $q$ represents the generalized coordinate(s). One can obtain the equations of motion by extremizing the action via $\delta S[q]/\delta q^k = 0$. The functional derivative is defined via
 \bq
\de S[q;\de q]=
\left.\frac{d S[q +\ep \de q]}{d \ep}\right|_{\ep =0}
=: \left\langle \frac{\de S[q]}{\de q^{i}},\de q^{i}\right\rangle\,.
\eq
The continuum version of the action principle is equally straightforward. We introduce a label $a$ that tracks a given fluid element; $q$ is now a function of $a$ and $t$. We introduce the deformation matrix, $\p q^i/\p a^j=:\q ij$ and the Jacobian, $\mathcal{J}:= \det(\q ij)$. The volume element for the fluid obeys
\bq
d^3q=\mathcal{J} d^3a\,,
\label{vol}
\eq 
and an area element is regulated via
\bq
(d^2q)_i=  \mathcal{J} \a ji \, (d^2a)_j\,,
\label{area}
\eq 
where $\mathcal{J} \a ji$ is defined to be the transpose of the cofactor matrix of $\q ji$. There are many identities that can be constructed involving $\q ij$ and $\calj$, but we desist from doing so; a detailed discussion is present in \cite{M98}.


\subsection{Attributes, observables and the Lagrange-Euler maps} 
\label{ssec:AOLE}
Our discussion has solely revolved around $q$, but a fluid element can also possess a certain density, entropy, etc. We refer to these quantities as {\it attributes},  since they are inherent to the fluid element. These quantities are dependent on $a$ alone and are Lagrangian constants of motion. The subscript $0$ is used to denote the attributes, to distinguish them from their Eulerian versions. 

The Eulerian version is commonly employed since it describes fields in terms of $r:=(x^1,x^2,x^3)$ and $t$, each of which can be tracked through experiments. Hence, we refer these fields as {\it observables}. However, a connection between the attributes and the observables, between the Lagrangian and Eulerian pictures is unclear. In order to transition back and forth, we must introduce maps that permit such an activity. We refer to these as the Euler-Lagrange or Lagrange-Euler maps, depending on the context. 

Before proceeding further, we remark that the model developed in this paper is 2D in nature, i.e. it has one ignorable coordinate. Hence, we shall treat our system as 3D in nature, but we note the existence of an implicit constraint $z = q_z(a,t) = a_z$, indicating that the $z$-component of the trajectory remains fixed.  

A natural starting point is the velocity field $v(r,t)$. In the Eulerian picture, we detect this velocity at a given instance in space and time. Intuitively, we expect the same result to hold true in the Lagrangian picture, i.e. the relation $\dot{q}(a,t)=v(r,t)$ must hold true. The LHS is the Lagrangian velocity, and the overdot indicates the derivative at fixed $a$. By the same logic, we expect the Eulerian and Lagrangian positions to coincide at this moment, ensuring that $r=q(a,t)$. We assume that this map is invertible, permitting us to obtain $a=q^{-1}(r,t)=:a(r,t)$. From the two conditions, we find that
\bq
v(r,t) =\left.\dot{q}(a,t)\right|_{a=a(r,t)}\,.
\eq
The above map constitutes a Lagrange to Euler map, as it expresses Eulerianizes the Lagrangian version. The property of invertibility also allows us to undertake the converse operation. 

Amongst the attributes transported by a fluid element, one of them is the entropy $s_0(a)$. The entropy is advected in an ideal fluid, implying that it is constant along a fluid trajectory. In other words, the Eulerian entropy $s(r,t)$ must equal the Lagranian entropy $s_0(a)$. Thus, we conclude that $s$ behaves as a zero form, or as a scalar. Next, we consider the density whose attribute--observable pair are denoted by $\rho_0(a)$ and $\rh(r,t)$. Our fluids must (typically) obey mass conservation, which is expressible as $\rho(r,t)d^3r=\rho_0d^3a$ in an infinitesimal volume. One can geometrically interpret this as $\rho$ behaving as a three form or as a scalar density. By using (\ref{vol}) we obtain $\rho_0=\rho \calj$. 

Hitherto, our attributes (and observables) have been purely thermodynamic. Let us consider the magnetic field and denote the attribute--observable pair by $B_0(a)$ and $B(r,t)$. At this point, we make an important observation regarding ideal MHD: it has the very special property that the flux is frozen. The frozen flux constraint is expressible as $B\cdot d^2r=B_0\cdot d^2a$, and from (\ref{area}) we obtain $\calj B^i=\q ij  \,B_0^j$.  In other words, the magnetic field $B$ can be interpreted as a two-form or as a vector density, but \emph{only} in the case of ideal MHD. In general, one cannot introduce two forms into extended MHD theories since the frozen flux condition is (apparently) not obeyed. In the next section, we shall tackle the issue of frozen flux, and present a new dynamical variable that, by construction, satisfies the frozen flux constraint.

In the preceding expressions, we evaluate the attributes at $a=q^{-1}(r,t)=:a(r,t)$, thereby completing the Lagrange-Euler maps. We can also express the above relations in an integral form. This is done by demanding that the Eulerian and Lagrangian observation points coincide, which is accomplished via the judicious use of the delta function $\delta(r-q(a,t))$. Before proceeding further, we remind the reader that our model is actually 2D in nature.

As an example, the relation for the density is shown
\bqy
\rh({r},t)&=&\int_D \!d^2a
\, \rh_0(a) \, \de\left({r}-{q}\left(a,t\right)\right)
\nonumber\\
&=&\left. \frac{\rh_0}{\mathcal{J}}\right|_{a=a({r},t)}\,.
\label{rhoEu3D}
\eqy
Instead of the velocity, we introduce the canonical momentum $M^c=\left(M^c_1,M^c_2\right)$, which is related to the Lagrangian canonical momentum as follows
\bqy 
M^c(r,t)&=&\int_D \!d^2a \,
{\Pi}(a,t) \, \de\left({r}-{q}(a,t)\right) 
\nonumber\\
&=& \left.
\frac{\Pi(a,t)}{\mathcal{J}}
\right|_{a=a(r,t)}\,.
\label{Mcan3D}
\eqy
In the case of ideal MHD,  $\Pi(a,t)=\left(\Pi_1,\Pi_2\right)=\rh_0 \dot{q}$.  One can always determine $\Pi(a,t)$ via $\Pi(a,t) = {\de L}/{\de \dot{q} }$, and use it in the above expression. In the case of gyroviscous fluids, one finds that $\Pi(a,t) \neq \rh_0 \dot{q}$ as shown in \cite{MLA14,LM14}.

\subsection{The Eulerian Closure Principle and action-building}
 \label{ssec:ECP}
 
Thus far, much of our analysis has been predicated on the notion that the Eulerian and Lagrangian pictures must be equivalent to one another, i.e.,  one can find a set of maps that allows us to go back and forth between the two descriptions. A natural consequence of this is that our action must also obey such a property.

This property is the ECP, referred to in the Introduction.  The ECP amounts to the following:  given an action expressed in terms of the Lagrangian variables, it must be equally expressible fully in terms of the Eulerian variables. For instance, consider the kinetic energy functional
\bq \label{KinLag}
\int \int \frac{1}{2} \rho_0 \dot{q}^2\,d^2a\, dt,
\eq
and let us invoke the Lagrange-Euler maps introduced in the previous subsection. Through suitable use of (\ref{rhoEu3D}) and (\ref{Mcan3D}), one can show that (\ref{KinLag}) reduces to
\bq \label{KinEul}
\int \int \frac{1}{2} \rho v^2\,d^2r\, dt.
\eq
As a result, we conclude that (\ref{KinLag}) satisfies the ECP since the kinetic energy functional in Lagrangian variables has been expressed in terms of Eulerian variables. The ECP dictates that all the other terms in the action also exhibit identical behavior as the kinetic energy functional. 

In summary, we follow a two step procedure to construct the action. The first involves the choice of the domain and the observables. The second involves the construction of each term in the action from first principles (when possible) while ensuring that they obey the ECP.

\section{The Inertial MHD action}
 \label{IMHDaction}
 
In this section, we shall present a new dynamical variable, one that determines a frozen flux for our model.  An action principle in terms of this new variable is developed, and the equations of motion are obtained and analyzed.

\subsection{The inertial magnetic field: a new dynamical variable}
 \label{ssec:inertialB}
 
In Section \ref{ssec:AOLE}, we discussed the implications of magnetic flux freezing in ideal MHD. Extended MHD lacks this feature, which implies that the magnetic field can no longer be interpreted as a two form. From a purely geometric point-of-view, it would be logical to look for a new dynamical variable, \emph{not} $B$, which could play a similar role. 

Hence, we introduce the variable $B_e$ and its corresponding attribute $B_{e0}$. The relation between the two is akin to that obeyed by the magnetic field in ideal MHD, viz. $\calj B^i_e=\q ij  \,B_{e0}^j$. Since we claim that our new theory is still a magnetofluid model, it is necessary for $B_e$ to be a function of the MHD variables $v$, $B$, $n$ and $s$. We make the choice
\bq \label{Bedef}
B_e = B + \frac{m_{e}}{e^{2}}\nabla\times\left(\frac{{ J}}{n}\right) = B + \frac{m_{e}}{\mu_0 e^{2}}\nabla\times\left(\frac{\left(\nabla \times B\right)}{n}\right).
\eq
In other words, this is also equivalent to stating that we replaced the vector potential $A$ by $A_e$, the latter of which is given by
\begin{equation} \label{Aedef}
A_e = A + \frac{m_{e}}{e^{2}}\left(\frac{{ J}}{n}\right) = A + \frac{m_{e}}{\mu_0 e^{2}}\left(\frac{\nabla \times B}{n}\right).
\end{equation}
Although the expressions (\ref{Bedef}) and (\ref{Aedef}) may appear \emph{ad hoc}, there are good reasons that justify these choices. The first stems from the inclusion of electron inertia, which is exemplified by the presence of an additional factor involving $m_e$ and it also satisfies the consistency requirement, i.e. in the limit $m_e/m_i \rightarrow 0$, we recover the usual magnetic field and vector potential. Secondly, we note that $B_e$ serves as a natural dynamical variable in extended MHD theories; for instance, if one takes the curl of equation (20) in \cite{KM14} and uses Faraday's law, we recover a dynamical equation for $B_e$. It is possible to carry out a similar procedure for the extended MHD models presented in \cite{S56,GP04} and arrive at the same conclusion. 

Lastly, the statement of flux freezing in ideal MHD is equivalent to stating that $\oint A \cdot dl$ is an invariant, which is altered  in our model.  To understand the alteration consider the canonical momentum for the electrons, which is proportional to $A - (m_e v_e/e)$. Assuming $v_e \gg v_i$  permits the approximation $J \approx -e n v_e$; consequently,   the canonical momentum is (approximately) equal to $A_e$. If we let $m_e/m_i \rightarrow 0$, the canonical momentum reduces to $A$. As this is a 2D theory, with $z$ serving as the ignorable coordinate, the corresponding canonical momentum in the $z$-direction is conserved. This yields  $\oint \left(A_e\right)_z  dz$, which is akin to $\oint A \cdot dl$ being conserved in ideal MHD. Later, we shall show that even better reasons can be advanced, albeit \emph{a posteriori}, that further justify the choice of $B_e$. 

Before we proceed to the next section, we introduce the nomenclature `inertial magnetic field' to refer to $B_e$. The choice is natural since $B_e$ plays the role of a magnetic field, and incorporates the effects of electron inertia. Hence, we refer to this theory as inertial MHD (IMHD).

\subsection{The IMHD action}
 \label{ssec:IMHDact}
 
We introduce the action for IMHD  below, and then comment on its significance and interpretation. Our variables are chosen to be the density (3-form) $\rho$, the inertial magnetic field (2-form) $B_e$, the entropy (0-form) $s$ and the velocity $v$.

\begin{equation} \label{IMHDact}
S = \int \int \left[\frac{\rh v^2}{2} - \rh U\left(\rh,s\right) - \frac{B_e \cdot B}{2\mu_0}\right]\,d^3r\, dt.
\end{equation}
The first term in (\ref{IMHDact}) is the kinetic energy, which was already shown to obey the ECP in Section \ref{ssec:ECP}. The second term in (\ref{IMHDact}) is the internal energy density, and is the product of density and the specific internal energy (per unit mass). This term generates the temperature and the pressure, given by $\partial U/\partial s$ and $\rh^2 \partial U/\partial \rh$ respectively. The third term in the above expression is the unusual part, since it deviates from the usual ideal MHD action. In the limit where $m_e/m_i \rightarrow 0$ we have noted that $B_e \rightarrow B$, which in turn reduces the last term of  (\ref{IMHDact}) to the conventional magnetic energy density. 

Although  (\ref{IMHDact}) is expressed in terms of the Eulerian variables, the ECP and the Euler-Lagrange maps allow us to express  (\ref{IMHDact}) purely in terms of the Lagrangian coordinate $q$ and the attributes. In order to do so, we express the magnetic field $B$ in terms of the inertial magnetic field $B_e$ as follows
\begin{equation}
B \left(r,t\right) = \int K\left(r|r'\right) B_e \left(r',t\right)\,d^2r',
\end{equation}
where $K$ is a complicated kernel. Using the kernel is quite complex, but we note that the self-adjoint property is preserved. Alternatively, one can use the Euler-Poincar\'e approach, described in Section \ref{Intro}, to obtain the same result. A short summary of which  is presented in Appendix \ref{AppA}.

Before proceeding to the next section, a couple of remarks regarding (\ref{IMHDact}) are in order. Firstly, the only term involving $\dot{q}$ is the kinetic energy term. Hence, one can perform a Legendre transformation, and recover the Hamiltonian (in Lagrangian variables). Upon Eulerianizing the Hamiltonian, we obtain
\begin{equation} \label{HamIMHD}
H = \int \left[\frac{\rh v^2}{2} + \rh U\left(\rh,s\right) + \frac{B_e \cdot B}{2\mu_0}\right]\,d^2r.
\end{equation}
We can use the definition of $B_e$, as given in (\ref{Bedef}), and simplify the resultant expression. The result is
\begin{equation} \label{EnergyIMHD}
H = \int \left[\frac{\rh v^2}{2} + \rh U\left(\rh,s\right) + \frac{B^2}{2\mu_0} + \frac{m_e}{n e^2}\frac{J^2}{2} \right]\,d^2r.
\end{equation}
The above expression is identical to equation (23) of \cite{KM14}. Furthermore, we see that (\ref{EnergyIMHD}) is identical to MHD Hamiltonian, except for the last term. As a consistency check, we verify that the last term does vanish in the limit $m_e/m_i \rightarrow 0$. These facts lend further credence to our choice of $B_e$ and the action (\ref{IMHDact}). 

\subsection{The IMHD equations of motion} \label{ssec:IMHDEOM}
The Lagrange-Euler maps outlined in Section \ref{ssec:AOLE} permit us to obtain the corresponding dynamical equations for the observables by applying $\partial/\partial t$ on both sides of the map. We obtain
\begin{equation} \label{sevol}
\frac{\partial s}{\partial t} + v\cdot \nabla s = 0,
\end{equation}
\begin{equation} \label{rhoevol}
\frac{\partial \rho}{\partial t} + \nabla \cdot \left(\rh v\right) = 0,
\end{equation}
\begin{equation} \label{Beevol}
\frac{\partial B_e}{\partial t} + B_e \left(\nabla \cdot v\right) - \left(B_e \cdot \nabla\right)v + \left(v \cdot \nabla\right) B_e = 0.
\end{equation}
The equations (\ref{sevol}), (\ref{rhoevol}) and (\ref{Beevol}) correspond to the Lie-dragging of zero, three and two forms respectively. From the definition of (\ref{Bedef}), we see that $\nabla \cdot B_e = 0$, and this implies that one can rewrite (\ref{Beevol}) as follows
\bqy \label{Beevolv2}
\frac{\partial B_e}{\partial t} &=& \nabla \times \left(v \times B_e\right) 
= \nabla \times \left(v \times B\right) \nonumber \\
&+& \frac{m_e}{ e^2}\nabla \times \left[v \times \left(\nabla \times \left(\frac{\nabla \times B}{n}\right) \right) \right].
\eqy
The equation of motion is obtained by extremizing the action in Lagrangian variables, or by extremizing the Eulerian action via the Euler-Poincar\'e approach. It is found to be
\bqy \label{vevol}
\rh \left(\frac{\partial v}{\partial t} + \left(v \cdot \nabla\right) v\right) &=& -\nabla p + J \times B \nonumber \\
&-& \frac{m_e}{e^2} \left(J \cdot \nabla\right)\left(\frac{J}{n}\right).
\eqy

Equations (\ref{Beevolv2}) and (\ref{vevol}) constitute the heart of our model. Let us first consider the latter. We see that it is nearly identical to the usual ideal MHD momentum equation, except for the presence of the last term, which can be neglected in the limit $m_e/m_i \rightarrow 0$.  However, this term represents more than a correction - in the extended MHD models with electron inertia, this term is absolutely crucial for energy conservation, as pointed out in \cite{KM14}. Secondly, we note that our equation of motion is exactly identical to equations (2) and (19) of \cite{KM14}, thereby lending further credence to our choice of the inertial magnetic field and action. 

We turn our attention to (\ref{Beevolv2}), which represents the extended Ohm's law. It is instructive to compare this against the inertial Ohm's law of \cite{KM14}, represented by their equation (20). We find that our expression is exactly identical to equation (20) of \cite{KM14}, when the 2D limit of their model is considered and $B_z \rightarrow \mathrm{const}$ (constant guide field) is assumed. Under these assumptions, the two results are exactly identical, irrespective of whether the fluid is compressible or incompressible. A few comments on the 3D generalization of this model are presented in Section \ref{ssec:BrackExt}.

To summarize thus far, we find that the momentum equations of our model and that of \cite{KM14} are identical. The Ohm's laws are also in perfect agreement with one another in the 2D, constant guide field limit. In addition, both of them yield the same (conserved) energies and momenta. Collectively, it is self-evident that these represent ample grounds for justifying the form of the inertial magnetic field $B_e$ and the IMHD action. 

\section{The Hamiltonian formulation of inertial MHD} \label{HamForm}
In this section, we describe the methodology employed in recovering the (Eulerian) noncanonical Hamiltonian picture from the (Lagrangian) canonical action. After obtaining the bracket--Hamiltonian pair, we comment on potential extensions of this framework.

\subsection{Derivation of the inertial MHD bracket} \label{ssec:IMHDbracket}

Our first step is the determination of the Hamiltonian, which is done via a Legendre transformation and Eulerianizing the resultant expression. The exercise was already performed in Section \ref{ssec:IMHDact}, and the Hamiltonian is given by (\ref{HamIMHD}). An alternative route is to invoke Noether's theorem, which also leads to the same result. 

Next, we need to obtain the noncanonical bracket. A detailed description of this procedure can be found in \cite{Mor09}; here, we shall merely present the salient details. Before proceeding on to the derivation, we reformulate our observables. We replace the velocity $v$ by the momentum $M^c$, and the entropy $s$ by the entropy density $\sigma = \rh s$. The new set of observables result in a simpler and compact noncanonical Poisson bracket. Let us recall from Section \ref{ssec:AOLE} that the Lagrange-Euler maps can be represented in an integral form. We present them below

\begin{equation} \label{rhointEL}
\rho=\int d^{2}a\,\delta\left(r-q(a,t)\right)\rho_{0}(a),
\end{equation}
\begin{equation}  \label{sigmaintEL}
\sigma=\int d^{2}a\,\delta\left(r-q(a,t)\right)\sigma_{0}(a),
\end{equation}
\begin{equation} \label{BeintEL}
B_e^{j}=\int d^{2}a\,\delta\left(r-q(a,t)\right)q_{,k}^{j}B_{e0}^{k}(a),
\end{equation}
\begin{equation} \label{MintEL}
M^{c}=\int d^{2}a\,\delta\left(r-q(a,t)\right)\Pi(a,t).
\end{equation}
The last expression is also equivalent to $M^c = \rh v$, which can be found by computing $\Pi$ from the Lagrangian, and then obtaining the Eulerian equivalent. We drop the subscript $c$ henceforth, since the canonical momentum $M^c$ is the same as the kinetic momentum $M = \rh v$.

Next, we note that a given functional can be expressed either in terms of the canonical momenta and coordinates, $\Pi$ and $q$, or in terms of the observables. Hence, we can denote the former by $\bar{F}$ and the latter by $F$, and note that $\bar{F}\equiv F$. As a result, we find that
\begin{eqnarray} \label{FuncEquv}
&&\int d^{2}a\,\frac{\delta\bar{F}}{\delta\Pi}\cdot\delta\Pi+\frac{\delta\bar{F}}{\delta q}\cdot\delta q \\
&&=\int d^{2}r\,\frac{\delta F}{\delta M^{c}}\cdot\delta M^{c}+\frac{\delta F}{\delta B}\cdot\delta B+\frac{\delta F}{\delta\rho}\delta\rho+\frac{\delta F}{\delta\sigma}\delta\sigma. \nonumber
\end{eqnarray}
From (\ref{rhointEL}), we can take the variation on the LHS and RHS, thereby obtanining
\begin{equation} \label{delrhovar}
\delta\rho=-\int d^{2}a\,\rho_{0}\nabla\delta\left(r-q(a,t)\right)\cdot\delta q.
\end{equation}
A similar procedure can also be undertaken for (\ref{sigmaintEL}), (\ref{BeintEL}) and (\ref{MintEL}) as well. We substitute (\ref{delrhovar}) into the second line of (\ref{FuncEquv}) and carry out an integration by
parts, and a subsequent interchange of the order of integration. This process is repeated for the rest of the variables. By doing so, we can determine the functional derivatives ${\delta\bar{F}}/{\delta q}$ and ${\delta\bar{F}}/{\delta\Pi}$ in terms of the functional derivatives of the observables. Next, we note that the canonical bracket is given by
\begin{equation} \label{CanBrack}
\{\bar{F},\bar{G}\} = \int d^2a\,\left(\frac{\delta\bar{F}}{\delta q} \cdot \frac{\delta\bar{G}}{\delta\Pi} - \frac{\delta\bar{G}}{\delta q} \cdot \frac{\delta\bar{F}}{\delta\Pi}\right).
\end{equation}
We can now substitute the expressions for ${\delta\bar{F}}/{\delta q}$ and ${\delta\bar{F}}/{\delta\Pi}$, obtained as per the procedure outlined above, into (\ref{CanBrack}) and derive the noncanonical bracket. It is found to be
\begin{align} \label{IMHDnoncanbrack}
\left\{ F,G\right\} = -\int d^{2}r\,&\Bigg[ M_{i}\left(\frac{\delta F}{\delta M_{j}}\partial_{j}\frac{\delta G}{\delta M_{i}}-\frac{\delta G}{\delta M_{j}}\partial_{j}\frac{\delta F}{\delta M_{i}}\right) \nonumber \\
&+ \rho\left(\frac{\delta F}{\delta M_{j}}\partial_{j}\frac{\delta G}{\delta\rho}-\frac{\delta G}{\delta M_{j}}\partial_{j}\frac{\delta F}{\delta\rho}\right) \nonumber \\
&+ \sigma\left(\frac{\delta F}{\delta M_{j}}\partial_{j}\frac{\delta G}{\delta\sigma}-\frac{\delta G}{\delta M_{j}}\partial_{j}\frac{\delta F}{\delta\sigma}\right) \nonumber \\
&+ B^{i}_e\left(\frac{\delta F}{\delta M_{j}}\partial_{j}\frac{\delta G}{\delta B^{i}_e}-\frac{\delta G}{\delta M_{j}}\partial_{j}\frac{\delta F}{\delta B^{i}_e}\right) \nonumber \\
&+ B^{i}_e\left(\frac{\delta G}{\delta B^{j}_e}\partial_{i}\frac{\delta F}{\delta M_{j}}-\frac{\delta F}{\delta B^{j}_e}\partial_{i}\frac{\delta G}{\delta M_{j}}\right)\Bigg].
\end{align}
The inertial MHD bracket, derived above, possesses a couple of remarkable features. Firstly, we note that the bracket is precisely identical to the ideal MHD bracket of \cite{MG80}, if we replace $B_e$ in (\ref{IMHDnoncanbrack}) with $B$ instead. Secondly, if one replaces $M$ by $M^c$ in the above expression, one can obtain an expression for the gyroviscous inertial MHD bracket, yielding results identical to those of \cite{Mor09,MLA14}. 

We must reiterate the importance of the bracket, because it further highlights the merits of $B_e$ as a dynamical variable. Our simple postulate in Section \ref{ssec:inertialB}, that $B_e$ behaves as a two form, ensures that inertial MHD and ideal MHD are identical to one another under the exchange $B_e \leftrightarrow B$. Not only does $B_e$ yield equations of motion that are highly similar to those of extended MHD, but it also maintains a close connection with ideal MHD via its notion of flux freezing. Owing to the near-identical nature of the inertial and ideal MHD brackets, an independent analysis of inertial MHD is not required; instead, one can simply migrate the results pertaining to the Casimirs, equilibria and stability of ideal MHD models, by replacing $B$ by $B_e$ in the suitable places. In particular, we note that the Casimir
\begin{equation}
\calc_1 = \int d^3r\, \rh f(s),
\end{equation}
still remains an invariant in inertial MHD. On the other hand, the counterpart of the magnetic helicity of ideal MHD is 
\begin{eqnarray}
\calc_2 &=& \int d^3r\, A_e \cdot B_e = \int d^3r\, \Bigg[A \cdot \left(\nabla \times A\right)  \\
&+& \frac{2m_e}{\mu_0 n e^2} B \cdot \left(\nabla \times B\right) + \frac{m_e^2}{e^4}\left(\frac{J}{n}\right)\cdot \left(\nabla \times \left(\frac{J}{n}\right)\right) \Bigg], \nonumber
\end{eqnarray}
and it is seen that each of the three terms is of the form $W \cdot \left(\nabla \times W\right)$. Note that the terms in the second line vanish when $m_e/m_i \rightarrow 0$, thereby yielding the ideal MHD magnetic helicity. The cross helicity of ideal MHD morphs into 
\begin{eqnarray}
\calc_3 &=& \int d^3r\, v \cdot B_e =  \int d^3r\, v \cdot \left(\nabla \times A_e\right)\\
&=& \int d^3r\, \left[v \cdot B + \frac{m_e}{e^2} v \cdot \left(\nabla \times \left(\frac{J}{n}\right) \right) \right], \nonumber
\end{eqnarray}
and we see that it reduces to the ideal MHD cross helicity if we assume $m_e/m_i \rightarrow 0$. It is easily seen that the ideal and inertial MHD cross helicities are both expressible as $v \cdot \left(\nabla \times W\right)$.

\subsection{The six-field model and its subcases}
 \label{ssec: 6fieldmodel}
 
Hitherto, we have not fully exploited the 2D nature of our model. The choice was deliberate since the bracket and the equations of motion could be expressed in a relatively compact form. However, it comes at the cost of obtaining a narrower class of Casimirs, and an inability to clearly demarcate the behavior of the different fields. We shall now exploit this 2D symmetry. 

First, let us consider $B_e$, defined via (\ref{Bedef}). We see that it is divergence free. As $z$ serves as our ignorable coordinate, we can immediately express it as
\begin{equation} \label{BeTS}
B_e = {B}_{ez} \hat{z} + \nabla \psi_e \times \hat{z}.
\end{equation}
Next, we consider the momentum, recognizing that it involves two components. Hence, the most general possible representation is
\begin{equation} \label{MTS}
M = \nabla \Gamma + \nabla \varphi \times \hat{z}.
\end{equation}

We note that a similar analysis, albeit in terms of $B$ instead of $B_e$, was carried out in \cite{AMP10,AMP12}. The advantage of inertial MHD is that the bracket is identical in structure to that of ideal MHD under the interchange $B_e \leftrightarrow B$. Upon substituting (\ref{BeTS}) and (\ref{MTS}) into (\ref{IMHDnoncanbrack}) and using the functional derivative chain rule, we obtain a bracket identical to that of equation (98) in \cite{AMP10}, except for two differences. The bracket obtained involves an integration over $d^2r$, as opposed to $d^3r$ in \cite{AMP10}. Secondly, one must set $M_z = 0$ in equation (98) in \cite{AMP10} as our model lacks the $z$-component of the velocity. 

In summary, we have a 6-field model with our observables given by $\left(\Gamma,\varphi,B_{ez},\psi_e,\rho,\sigma\right)$. Each of these six fields are now scalar and possess clear physical interpretations. We can whittle the model down to a 5-field model by assuming it to be isentropic, which eliminates $\sigma$. If we assume incompressibility, we eliminate $\rho$ and $\Gamma$ - the last of which follows from the condition $\nabla \cdot v = 0$. Lastly, we can eliminate the guide field $B_{ez}$ by making it constant, and our resultant model now involves just $\varphi$ and $\psi_e$. We introduce the notation $\omega = \Delta \varphi$, implying that the two functional derivatives are related via $\Delta F_\omega = - F_\varphi$, and the final bracket is given by
\begin{align} \label{OPBrack}
\{F,G\}=-\int d^2r\,&\Big[\omega[F_{\omega},G_{\omega}]+\psi_e\Big([F_{\omega},G_{\psi_e}] \nonumber \\
&-[G_{\omega},F_{\psi_e}]\Big)\Big],
\end{align}
and the corresponding Hamiltonian takes on the form
\begin{equation} \label{OPHam}
H=\int d^2r \frac{1}{2}\left[d_e^2 (\nabla^2\psi)^2+\frac{|\nabla\psi|^2}{\mu_0}+\frac{|\nabla\varphi|^2}{\rho}\right],
\end{equation}
where $B = \nabla \psi \times \hat{z}$, and the relation between $\psi_e$ and $\psi$ is determined via (\ref{Bedef}). We note that $d_e$ represents the ion skin depth. The bracket and Hamiltonian, given by (\ref{OPBrack}) and (\ref{OPHam}) are of considerable importance, as they give rise to the well known Ottaviani-Porcelli model \cite{OP93}, used in  modelling collisionless magnetic reconnection. 

\subsection{Extensions of the inertial MHD bracket}
 \label{ssec:BrackExt}
 
In the preceding subsection, we have obtained the inertial MHD noncanonical bracket, with the corresponding expression given by (\ref{IMHDnoncanbrack}). A crucial feature of inertial MHD was also identified, namely, the close affinity with the ideal MHD bracket, as one can be transformed into the other via $B_e \leftrightarrow B$.

The analogy between $B$ and $B_e$ also makes it possible to import the results of 2D gyroviscous MHD, and recast them in an inertial MHD framework. As noted in the previous subsection, the noncanonical brackets derived in \cite{MLA14} can be adapted for such a purpose. They are easily distinguishable from the non-gyroviscous brackets owing to the presence of the canonical momentum $M^c$ in place of the kinetic momentum $M$. If the same methodology is employed herein, we can obtain a model for 2D gyroviscous inertial MHD. It must be cautioned, however, that these methods are only applicable to the inertial and ideal brackets, as they are equivalent under $B_e \leftrightarrow B$. Modifying the Hamiltonians is a trickier task, as it requires us to explicitly use the relation (\ref{Bedef}).

As we have stated thus far, our model of inertial MHD possesses an ignorable coordinate, thereby rendering it 2D. A natural generalization of the procedure is to undertake the same work in a 3D framework. Our central results thus far were the equation of motion (\ref{vevol}) and the Ohm's law (\ref{Beevolv2}). We find that the former is unmodified, and is identical to that of \cite{KM14}. However, in the 3D limit, we find that the Ohm's law of \cite{KM14} and our model are \emph{not} in agreement, although most of the terms are identical to one another. Next, we consider the incompressible 3D limit, and we find that there is a near-exact match; in fact, we find that the Ohm's law of \cite{KM14} reduces to our Ohm's law (\ref{Beevolv2}) when the flow is irrotational.

\section{Conclusion}
 \label{Conc}
 
In this paper, we have approached the issue of electron inertia in an unusual manner - via the inclusion of geometric constraints. We generalized the flux freezing condition of ideal MHD, by replacing the vector potential $A$ with an extended vector potential $A_e$, and motivated it via the conservation of canonical momentum. Our model, dubbed 2D inertial MHD, comes with an intrinsic advantage - it is endowed with flux conservation, albeit not for the magnetic field. 

2D inertial MHD was shown to possess a couple of pleasing properties. Firstly, it yielded an equation of motion and an Ohm's law that were identical to the ones derived in \cite{KM14,KLMWW14}, when the 2D case of the latter, in the constant guide field limit, was taken. Secondly, we demonstrated that 2D inertial MHD could be expressed as a six (scalar) field model. A limiting subcase of 2D inertial MHD was shown to reproduce the   Ottaviani-Porcelli model \cite{OP93} of magnetic reconnection. Lastly, we demonstrated the inertial MHD bracket was identical to that of ideal MHD under the interchange $B_e \leftrightarrow B$, thereby cementing the close connection between the two models. 

There are several avenues that open up for investigation. It is possible, akin to ideal MHD, to derive expressions for compressible waves and nonlinear Alfv\'en-like solutions for inertial MHD. A second possibility is to move to the weak 3D limit, and obtain a suitable extension of the Ottaviani-Porcelli. We expect to tackle such issues in our subsequent work. 

\appendix
\section{The Euler-Poincar\'e approach to magnetofluids}
 \label{AppA}
 
Descriptions of the Euler-Poincar\'e formalism can be found in \cite{FR60,HMR98a,LM14}. Let us represent the Lagrangian density given in  (\ref{IMHDact}) by $\call$. The equation of motion, via the Euler-Poincar\'e approach is
\begin{equation} \label{EPEOM}
\frac{\partial M^c_i}{\partial t} + \frac{\partial T^j_i}{\partial x^j} = 0,
\end{equation}
where $M^c$ is the canonical momentum introduced in (\ref{Mcan3D}), and is given by $M^c_i = \partial \call/\partial v^i$. The stress tensor in the above equation is given by
\begin{eqnarray} \label{StressT}
T^j_i &=& M^c_i v^j + \frac{\partial \call}{\partial B_e^i}B_e^j \nonumber \\
&+& \delta^j_i \left(\call - \rh \frac{\partial \call}{\partial \rh} - B_e^k \frac{\partial \call}{\partial B_e^k}\right).
\end{eqnarray}
Note that the above expression is independent of the variable $s$. Upon taking the functional derivatives of $S$ with respect to the given variables, and substituting them into (\ref{StressT}) and (\ref{EPEOM}), we obtain the equation of motion (\ref{vevol}).

\bibliographystyle{apsrev4-1} 

%

\end{document}